\documentclass{osa-article}

\journal{osac}

\articletype{Research Article}

\begin{document}

\title{Large and robust mechanical squeezing of optomechanical systems in a highly
unresolved sideband regime via Duffing nonlinearity and intracavity
squeezed light}

\author{Jian-Song Zhang \authormark{1} and Ai-Xi Chen\authormark{2,3,*}}

\address{\authormark{1}Department of Applied Physics, East China Jiaotong University,
Nanchang 330013, People's Republic of China\\
\authormark{2}Department of Physics, Zhejiang Sci-Tech University, Hangzhou 310018, People's Republic of China\\
\authormark{3}Institute for Quantum Computing, University of Waterloo, Ontario N2L3G1, Canada}

\email{\authormark{*}aixichen@zstu.edu.cn} 



\begin{abstract}
We propose a scheme to generate strong and robust mechanical squeezing in an optomechanical system in the highly unresolved sideband (HURSB) regime
with the help of the Duffing nonlinearity and intracavity squeezed light. The system is formed by a standard optomechanical system with
the Duffing nonlinearity (mechanical nonlinearity) and a second-order nonlinear medium (optical nonlinearity).
In the resolved sideband regime, the second-order nonlinear
medium may play a destructive role in the generation of mechanical squeezing.
However, it can significantly increase the mechanical squeezing (larger than 3dB)
in the HURSB regime when the parameters are chosen appropriately. 
Finally, we show the mechanical squeezing is robust against the thermal fluctuations of the mechanical resonator.
The generation of large and robust mechanical squeezing in the HURSB regime is a combined effect of the mechanical and optical
nonlinearities.
\end{abstract}

\section{Introduction}
Optomechanical systems have received a lot of attentions
due to the wide range of applications such as highly sensitive measurement of tiny displacement
and quantum information processing \cite{Aspelmeyer2014,Bowen2015}.
In the highly sensitive measurement of tiny displacement, quantum squeezing of mechanical mode
is indispensable. In principle, quantum squeezing can be accomplished by the parametric interaction
of a quantum system \cite{Walls2008}. However, quantum
squeezing in this scheme can not be larger than 3dB since a quantum system becomes unstable
if the quantum squeezing is larger than 3dB as pointed out by Milburn and Walls \cite{Milburn1981}.

In recent years, many schemes have been proposed to generate strong mechanical squeezing beyond the 3dB limit including
continuous weak measurement and feedback \cite{Ruskov2005,Clerk2008,Szorkovszky2011,Szorkovszky2013},
squeezed light \cite{Jahne2009,Huang2010},
quantum-reservoir engineering \cite{Rabl2004,Zhang2009,Gu2013,Wangdongyang20191,Wangdongyang20192,Kronwald2013,Wollman2015,Pirkkalainen2015,Lecocq2015,Lei2016},
strong intrinsic nonlinearity \cite{Asjad2014,Lv2015}, and frequency modulation \cite{Mari2009,Han2019,Bai2020}.
For instance, large steady-state mechanical
squeezing can be achieved by applying two driving lasers to a cavity in
an optomechanical system \cite{Kronwald2013}. In this scheme, the power of the red-detuned driving field should be
larger than that of the blue-detuned driving field.
This scheme was realized experimentally in 2015 \cite{Wollman2015}. 
Very recently, the authors of \cite{Bai2020} have shown that larger mechanical squeezing can also be achieved with only one periodically amplitude-modulated
external driving field.

It is worth noting that the above schemes \cite{Ruskov2005,Clerk2008,Szorkovszky2011,Szorkovszky2013,Jahne2009,Huang2010,Rabl2004,Mari2009,Zhang2009,Gu2013,Wangdongyang20191,Wangdongyang20192,Kronwald2013,Wollman2015,Asjad2014,Lv2015,Bai2020}
are not valid to realize large mechanical squeezing beyond
the 3dB limit in the HURSB regime with $\kappa \gg \omega_m$. Here, $\kappa$ is the decay rate of the cavity and $\omega_m$ is the frequency of the mechanical resonator.
In order to generate large mechanical squeezing beyond the resolved sideband regime, the authors of \cite{Han2019} suggested to use
frequency modulation acting on both the cavity field and mechanical resonator. They have shown that mechanical squeezing beyond 3dB can be achieved in the
presence of frequency modulation beyond the resolved sideband and weak-coupling limits. It was shown that the strong mechanical squeezing beyond 3dB in the
unresolved sideband regime ($\kappa \approx 30 \omega_m$) can also be achieved by adding two auxiliary cavities
since the unwanted counter-rotating terms could be suppressed significantly with the help of quantum interference
from the auxiliary cavities \cite{ZhangRong2019}. However, the decay rates of the auxiliary cavities must be much smaller than the frequency of the mechanical resonator in the above scheme.
Later, we proposed a scheme to generate large mechanical squeezing beyond 3dB in the HURSB regime by adding two two-level atomic ensembles
and two driving lasers with different amplitudes \cite{Zhang2020}.
Very recently, it was shown that the quantum ground-state cooling of mechanical resonator in an optomechanical system
can be accomplished using intracavity squeezed light produced by a second-order nonlinear medium in the optomechanical system\cite{Clark2017,Lau2019,Asjad2019,Gan2019}.

In the present work, we propose a scheme to generate large and robust mechanical squeezing in the HURSB regime via the Duffing nonlinearity of the mechanical mode
and a second-order nonlinear medium in the cavity. The mechanical squeezing of the mechanical resonator can be larger than 3dB and is robust against the
thermal fluctuations of the mechanical resonator. This is a combined effect of nonlinearity-induced
parametric amplification (Duffing nonlinearity) and quantum ground-state cooling of the optomechanical system (intracavity squeezed light).
On the one hand, in the resolved sideband regime, the
second-order nonlinear medium may decrease the mechanical squeezing. On the other hand, the second-order nonlinear medium can significantly increase
the mechanical squeezing in the HURSB regime for realistic parameters when they are chosen appropriately.

\section{Model and Hamiltonian}
\begin{figure}[tbp]
\centering {\scalebox{1}[1]{\includegraphics{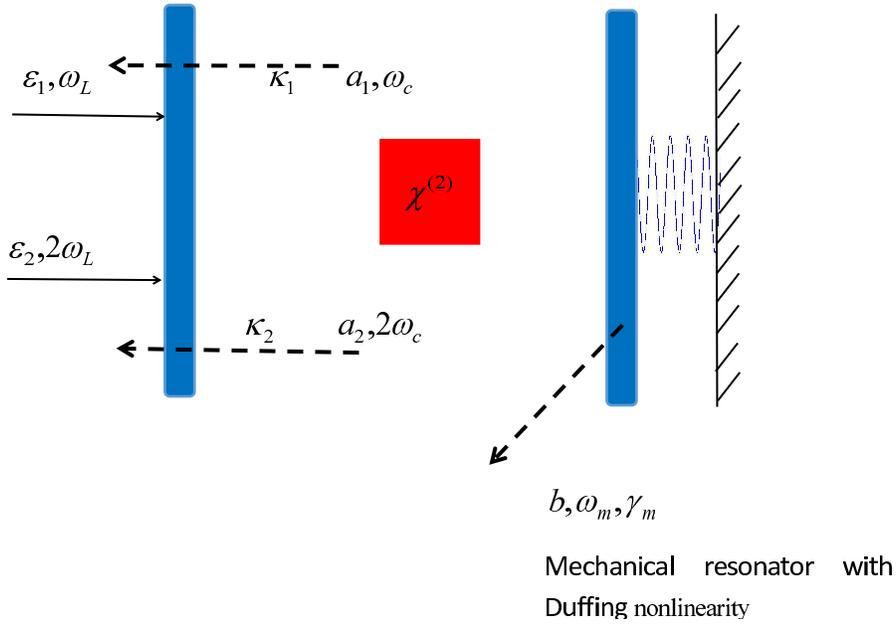}}}
\caption{Schematic representation of our model. The movable mirror is perfectly reflecting.
However, the fixed mirror is partially transmitting.
A second-order nonlinear medium denoted by $\chi^{(2)}$ is put into the cavity.
The fundamental and second-order optical modes with frequencies $\omega_c$ and $2\omega_c$ are denoted by $a_1$ and $a_2$.
Here, $\kappa_1$ and $\kappa_2$ are the decay rates of the fundamental and second-order optical modes, respectively.
The mechanical resonator (movable mirror) with frequency $\omega_m$ and decay rate $\gamma_m$ is denoted by $b$.
} \label{fig1}
\end{figure}

In the present work, we consider an optomechanical system formed by two mirrors as shown in Fig.1. One mirror is fixed and partially transmitting.
The other mirror is movable and perfectly reflecting. In addition, a second-order nonlinear medium $\chi^{(2)}$ is put into the Fabry-Perot cavity.
The fundamental mode and second-order optical mode are represented by $a_1$ and $a_2$ with frequencies $\omega_c$ and $2\omega_c$, respectively. The decay rates of
the two optical modes are $\kappa_1$ and $\kappa_2$. The movable mirror (mechanical oscillator) is denoted by $b$ with frequency $\omega_m$ and decay rate
$\gamma_m$. In addition, two driving fields with amplitudes $\varepsilon_1$ and $\varepsilon_2$ are applied to fundamental and second-order modes.
The Hamiltonian of the present model is (we set $\hbar = 1$)
\begin{eqnarray}
H &=& H_0 + H_{dr} + H_I + H_{D} + H_{N},\\
H_0 &=& \omega_c a_1^{\dagger}a_1 + 2\omega_c a_2^{\dag}a_2 + \omega_m b^{\dagger}b, \\
H_{dr} &=& i(\varepsilon_1 e^{-i\omega_L t}a_1^{\dagger} + \varepsilon_2 e^{-2i\omega_L t}a_2^{\dagger} - H.c.), \\
H_I &=& - g_1a_1^{\dagger}a_1(b^{\dagger} + b) - g_2a_2^{\dagger}a_2(b^{\dagger} + b),\\
H_{D} &=& \frac{\eta}{2}(b^{\dagger} + b)^4, \\
H_N &=&  \frac{i \chi_0}{2} (a_1^{\dagger 2} a_2 - a_1^2a_2^{\dagger}),
\end{eqnarray}
where $H_0$ is the free Hamiltonian of the whole system. $H_{dr}$ is the Hamiltonian for driving fields applied to
the fundamental and second-order modes with frequencies $\omega_L$ and $2\omega_L$.
The amplitudes of two driving fields are $\varepsilon_1 = \sqrt{2\kappa_1 P_1/\omega_L}$ and $\varepsilon_2 = \sqrt{2\kappa_2 P_2/(2\omega_L)} $.
$H_I$ is the interaction between the optical and mechanical modes.
The coupling strength between the mechanical mode and fundamental mode (second-order mode) is denoted by $g_1$ ($g_2$).
The Duffing nonlinearity of the mechanical mode is represented by $H_D$. It was pointed out that a nonlinear amplitude of
$\eta = 10^{-4}\omega_m$ can be achieved by coupling the mechanical mode to an auxiliary system \cite{Lv2015}.
The Hamiltonian of a second-order nonlinear medium is denoted by $H_N$ with $\chi_0$ being the interaction between the fundamental and second-order optical modes \cite{Lau2019,Asjad2019,Gan2019}.

In a rotating frame defined by the unitary transformation $U(t) = \exp{ \{-i\omega_L t(a_1^{\dagger}a_1 + 2 a_2^{\dagger}a_2)\} }$, we obtain the Hamiltonian as follows
\begin{eqnarray}
H &=& U^{\dagger}HU - iU^{\dagger}\dot{U} \nonumber\\
&=& \overline{\Delta}_c a_1^{\dagger}a_1 + 2\overline{\Delta}_c a_2^{\dagger}a_2 + \omega_m b^{\dagger}b \nonumber \\
&& + i(\varepsilon_1 a_1^{\dagger} + \varepsilon_2 a_2^{\dagger} - \varepsilon_1 a_1 - \varepsilon_2 a_2) \nonumber\\
&& - (g_1a_1^{\dagger}a_1 + g_2a_2^{\dagger}a_2)(b^{\dagger} + b) \nonumber\\
&& + \frac{\eta}{2}(b^{\dagger} + b)^4 + \frac{i \chi_0}{2} (a_1^{\dagger 2} a_2 - a_1^2a_2^{\dagger}),
\end{eqnarray}
with $\overline{\Delta}_c = \omega_c - \omega_L$.

\section{Quantum Langevin equations }

\begin{figure}[tbp]
\centering {\scalebox{0.5}[0.5]{\includegraphics{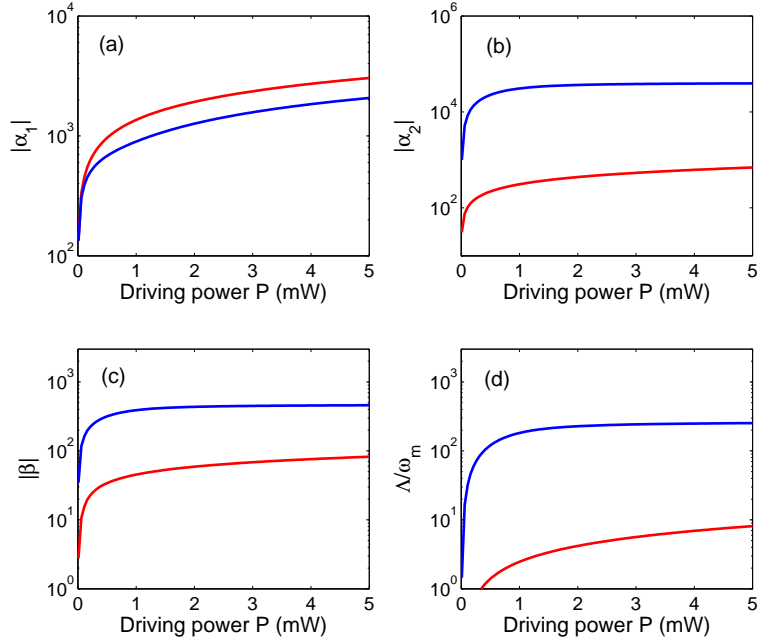}}}
\caption{ In Figs.2(a)-(c), the steady state amplitudes $|\alpha_1|$ , $|\alpha_2|$, and $|\beta|$ are plotted as functions of the driving power P for $|\chi_0| = 0$ (red lines) and $|\chi_0| = 10^{-3}\omega_m$ (blue lines). In Fig.2(d), the parameter $\Lambda$ is plotted as a function of the driving power P for $|\chi_0| = 0$ (red line) and $|\chi_0| = 10^{-3}\omega_m$ (blue line). We assume $P_1 = P_2 = P$ in the present work. 
Other parameter values are $\omega_m = 2\pi \times 20 MHz$, $\omega_L = 2\pi \times 500  THz$, $\eta = 10^{-4}\omega_m$, $g_1 = g_2 = g_0 = 10^{-4}\omega_m$, $\kappa_1 = 100\omega_m$, $\kappa_2 = 2000\omega_m$, $\gamma_m = 10^{-6}\omega_m$, $\varphi_b = n \pi$, $\Delta_c = 10\omega_m$, and $\Delta_c' = 20\omega_m$.
} \label{fig2}
\end{figure}

First, we linearize the above Hamiltonian by employing the following displacement transformations
$a_1 \rightarrow \alpha_1 + \delta a_1$, $a_2 \rightarrow \alpha_2 + \delta a_2$, and $b \rightarrow \beta + \delta b$.
The quantum Langevin equations are
\begin{eqnarray}
\dot{\alpha}_1 &=& -(i\Delta_c + \frac{\kappa_1}{2})\alpha_1 + \chi_0 \alpha_1^*\alpha_2 + \varepsilon_1,\nonumber \\
\dot{\alpha}_2 &=& -(i\Delta'_c + \frac{\kappa_2}{2})\alpha_2 - \frac{\chi_0}{2} \alpha_1^2 + \varepsilon_2,\nonumber \\
\dot{\beta} &=& -(i\omega_m + \frac{\gamma_m}{2})\beta -i\eta(16|\beta|^3\cos^3{\varphi_b} + 12 |\beta|\cos{\varphi_b})  
+ i g_1|\alpha_1|^2 + i g_2|\alpha_2|^2, \nonumber\\
\delta \dot{a}_1 &=& -(i\Delta_c + \frac{\kappa_1}{2})\delta a_1 + i G_1 (\delta b^{\dagger} + \delta b) \nonumber\\
&&+ \chi_0\alpha_2 \delta a_1^{\dagger} + \chi_0\alpha_1^*\delta a_2 + \sqrt{\kappa_1} a_{1,in}, \\
\delta \dot{a}_2 &=& -(i\Delta'_c + \frac{\kappa_2}{2})\delta a_1 + i G_2 (\delta b^{\dagger} + \delta b) - \chi_0\alpha_1 \delta a_1 + \sqrt{\kappa_2} a_{2,in}, \nonumber \\
\delta \dot{b} &=& -(i\omega_m + \frac{\gamma_m}{2})\delta b - 2i\Lambda (\delta b^{\dagger} + \delta b) \nonumber\\
&& + i(G_1 \delta a_1^{\dagger} + G_1^*\delta a_1) + i(G_2 \delta a_2^{\dagger} + G_2^*\delta a_2) + \sqrt{\gamma_m} b_{in},\nonumber
\end{eqnarray}
where $\Delta_c = \overline{\Delta}_c - 2 g_1|\beta|\cos{\varphi_b}$, $\Delta'_c = 2\overline{\Delta}_c - 2 g_2|\beta|\cos{\varphi_b}$, $\Lambda = 3\eta(4|\beta|^2\cos^2{\varphi_b} + 1)$,
and $G_{1,2} = g_{1,2}\alpha_{1,2}$. Here, we have assumed $\beta = |\beta| e^{i\varphi_b}$
and the amplitudes of two driving fields are $\varepsilon_1 = \sqrt{2\kappa_1 P_1/\omega_L}$ and $\varepsilon_2 = \sqrt{2\kappa_2 P_2/(2\omega_L)} $.

In Figs.2(a)-(d), we plot the parameters $|\alpha_1|$ , $|\alpha_2|$, $|\beta|$, and $\Lambda$ as functions of the driving power $P$ for $|\chi_0| = 0$ (red lines) and $|\chi_0| = 10^{-3}\omega_m$ (blue lines) using Eqs.(8). The stability condition is satisfied according to the Routh-Hurwitz criterion \cite{Dejesus1987}. In the presence of the nonlinear interaction between two cavity modes, we get $|\alpha_1| \approx 2000$, $|\alpha_2| \approx 4\times 10^4$, $|\beta| \approx 500$, and $\Lambda \approx 300\omega_m$ when the driving power $P$ is about 5 mW. Comparing the blue line with the red line of Fig.2(d), we find the parameter $\Lambda$ can be significantly enhanced by the nonlinear interaction between two cavity modes $\chi$. In the present work, we assume $|\chi| = 0.4\kappa_1$.
In the case of $\kappa_1 = 100\omega_m$, we find 
$|\chi| = 40\omega_m$. From the definition $|\chi| = |\chi_0| |\alpha_2|$, we observe that $|\chi| \approx 40\omega_m$ is possible since $\chi_0 = 10^{-3}\omega_m$ and $|\alpha_2|$ can be $4\times 10^4$.
In addition, $G_1 = g_1 |\alpha_1| \approx 0.2\omega_m$ when $g_1 = 10^{-4}\omega_m$ and $|\alpha_1| \approx 2000$. Therefor, the parameters $\chi$ and $G_1$ can be consistent with the value of $\Lambda$.

In the limit of large $\kappa_2$, the fluctuations of mode $a_2$ can be neglected and the adiabatic approximation is valid \cite{Asjad2019}.
Thus, the quantum Langevin equations can be reduced to
\begin{eqnarray}
\delta \dot{a}_1 &=& -(i\Delta_c + \frac{\kappa_1}{2})\delta a_1 + i G_1 (\delta b^{\dagger} + \delta b) + \chi \delta a^{\dagger}_1 + \sqrt{\kappa_1} a_{1,in} \nonumber\\
\delta \dot{b} &=& -(i\omega_m + \frac{\gamma_m}{2})\delta b - 2i\Lambda (\delta b^{\dagger} + \delta b) \nonumber \\
&& + iG_1 (\delta a_1^{\dagger} + \delta a_1) + \sqrt{\gamma_m} b_{in},
\end{eqnarray}
where $\chi = \chi_0 \alpha_2 = |\chi|e^{2i\phi}$. Without loss of generality, $G_1$ has been assumed to be real.

Now, we define the following quadrature operators $ X_{O = a_1,b} = (\delta O^{\dag} + \delta O)/\sqrt{2}$
and $ Y_{O = a_1,b} = i(\delta O^{\dag} - \delta O)/\sqrt{2}$.
The noise quadrature operators are defined as $ X^{in}_{O = a_1,b} = (O_{in}^{\dag} + O_{in})/\sqrt{2}$
and $ Y^{in}_{O = a_1,b} = i(O_{in}^{\dag} - O_{in})/\sqrt{2}$. From the above quantum Langevin equations, we obtain
\begin{eqnarray}
\dot{\vec{f}} &=& A \vec{f} + \vec{n}, \label{dfdt}
\end{eqnarray}
where $\vec{f} = ( X_{a_1}, Y_{a_1}, X_b, Y_b)^T$ and

\begin{eqnarray}
\vec{n} &=& (\sqrt{\kappa_1} X_{a_1}^{in}, \sqrt{\kappa_1} Y_{a_1}^{in}, \sqrt{\gamma_m} X_b^{in}, \sqrt{\gamma_m} Y_b^{in})^T,\\
A &=&\left(
\begin{array}{cccc}
|\chi|\cos{(2\phi)} - \frac{\kappa_1}{2} & |\chi|\sin{(2\phi)} + \Delta_c & 0 & 0\\
|\chi|\sin{(2\phi)} - \Delta_c           & -|\chi|\cos{(2\phi)} - \frac{\kappa_1}{2} & 2G_1 & 0\\
0 & 0 & -\frac{\gamma_m}{2} & \omega_m \\
2G_1 & 0 & -\omega_m - 4\Lambda & -\frac{\gamma_m}{2}
\end{array}  \right). \label{A}
\end{eqnarray}

Note that the dynamics of the present system described by Eq. (\ref{dfdt}) can be completely described by a $4\times4$ covariance matrix
$V$ with $V_{jk} = \langle f_j f_k + f_k f_j\rangle/2$. Using the definitions of $V$, $\vec{f}$,
and the above equations, we obtain the evolution of the covariance matrix $V$ as follows
\begin{eqnarray}
\dot{V} = A V + V A^T + D, \label{dV}
\end{eqnarray}
where $D$ is the noise correlation defined by
$D = diag[\frac{\kappa_1}{2}, \frac{\kappa_1}{2}, \frac{\gamma_m}{2}(2 n_{th} + 1), \frac{\gamma_m}{2}(2 n_{th} + 1)]$.
Here, $n_{th}$ is the mean phonon number of the mechanical resonator.

\section{Large and robust mechanical squeezing beyond resolved sideband regime}
\subsection{Stability}

\begin{figure}[tbp]
\centering {\scalebox{0.6}[0.6]{\includegraphics{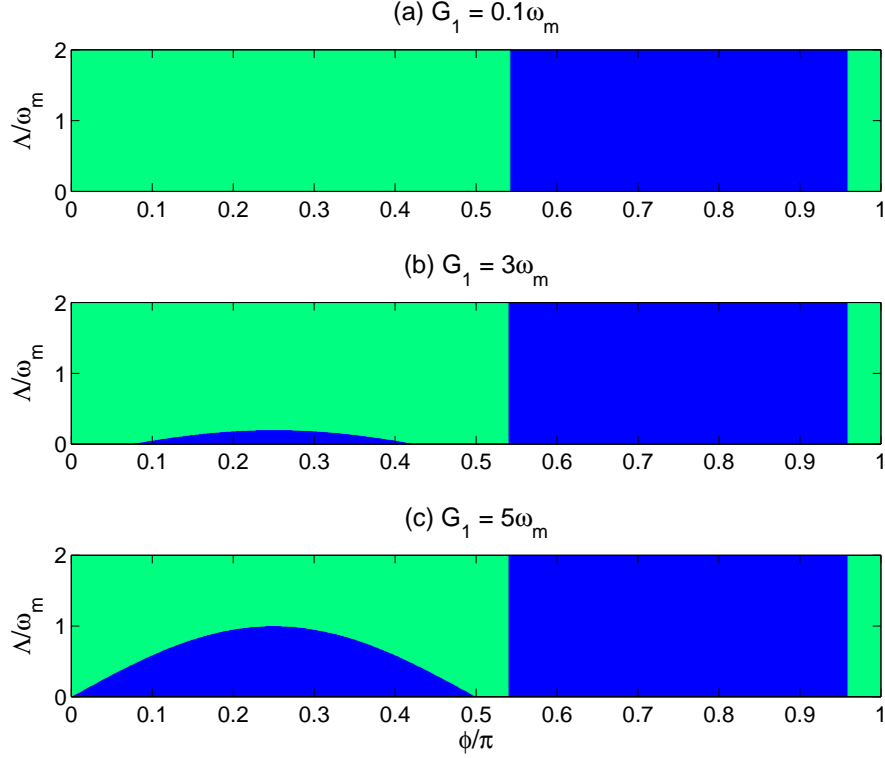}}}
\caption{ Stability of the present model. Here, $\chi = \chi_0 \alpha_2 = |\chi|e^{2i\phi}$ and $\Lambda = 3\eta(4|\beta|^2\cos^2{\varphi_b} + 1)$
are directly related to the second-order nonlinear medium and the Duffing nonlinearity of the mechanical mode.
We consider the HURSB case with $\kappa_1 \gg \omega_m$.
The stable and unstable regimes are represented by the green and blue regions, respectively.
Other parameter values are $\gamma_m = 10^{-6}\omega_m$, $\Delta_c = 10\omega_m$, $\kappa_1 = 100\omega_m$, $n_{th} = 1000$, $\varphi_b = n\pi (n=0,1,2,...)$, and $|\chi| = 0.4\kappa_1$.
} \label{fig3}
\end{figure}

We first investigate influence of the Duffing nonlinearity and optomechanical coupling strength on the
the stability of the present system. The Duffing nonlinearity and optomechanical coupling constant are related
to parameters $\Lambda = 3\eta(4|\beta|^2\cos^2{\varphi_b} + 1)$ and $G_1 = g_1 \alpha_1$, respectively.
It is well known that the system described by Eq. (\ref{dV}) is
stable only if all the real parts of the eigenvalues of the matrix A are negative according to the Routh-Hurwitz criterion \cite{Dejesus1987}.

From Fig.\ref{fig3}, we find the system is unstable for $0.54 < \phi < 0.95$. Comparing three panels of this figure, one can see
that the areas of the unstable regions could increase with the increase of the coupling constant $G_1$.
For example, the system is stable with $\phi = 0.2$, $\Lambda = \omega_m$ and $G_1 = 0.1\omega_m$ as one can find in the upper panel of Fig.\ref{fig3}.
However, if we increase the effective optomechanical coupling strength $G_1$ from $0.1\omega_m$ to $5 \omega_m$ the system is not stable with $\Lambda \leq \omega_m$ and $\phi = 0.2$.
Fortunately, the system can be stable if the Duffing nonlinearity is considered with $\Lambda > \omega_m$ (see the lowest panel of this figure).
Thus, the present system could be stable even for the strong and deep-strong coupling cases in the presence of the Duffing nonlinearity
which is consistent with the results of \cite{Lv2015}.

\subsection{Mechanical squeezing in HURSB regime}

\begin{figure}[tbp]
\centering {\scalebox{0.4}[0.4]{\includegraphics{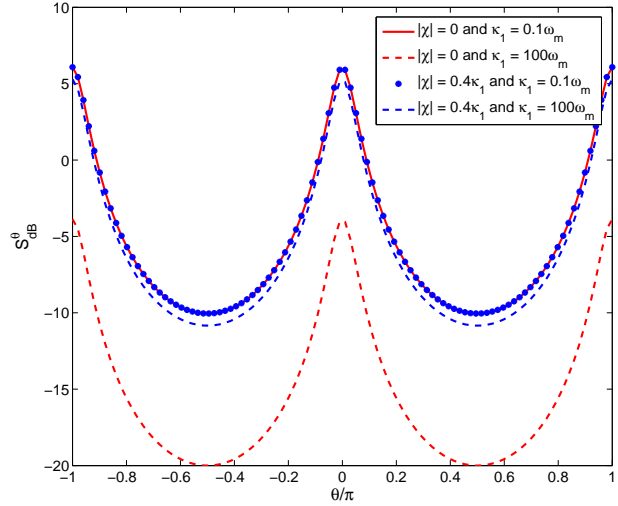}}}
\caption{ The mechanical squeezing $S^{\theta}_{dB}$ are plotted as functions of $\theta$ for different values of $|\chi|$ and $\kappa_1$. Other parameter values are $\gamma_m = 10^{-6}\omega_m$, $n_{th} = 0$, $G_1 = 0.1\omega_m$, $\phi = 0.5\pi$, $\varphi_b = n\pi$, $\Delta_c = 10\omega_m$, and $\Lambda = 10\omega_m$.
} \label{fig4}
\end{figure}

\begin{figure}[tbp]
\centering {\scalebox{0.6}[0.6]{\includegraphics{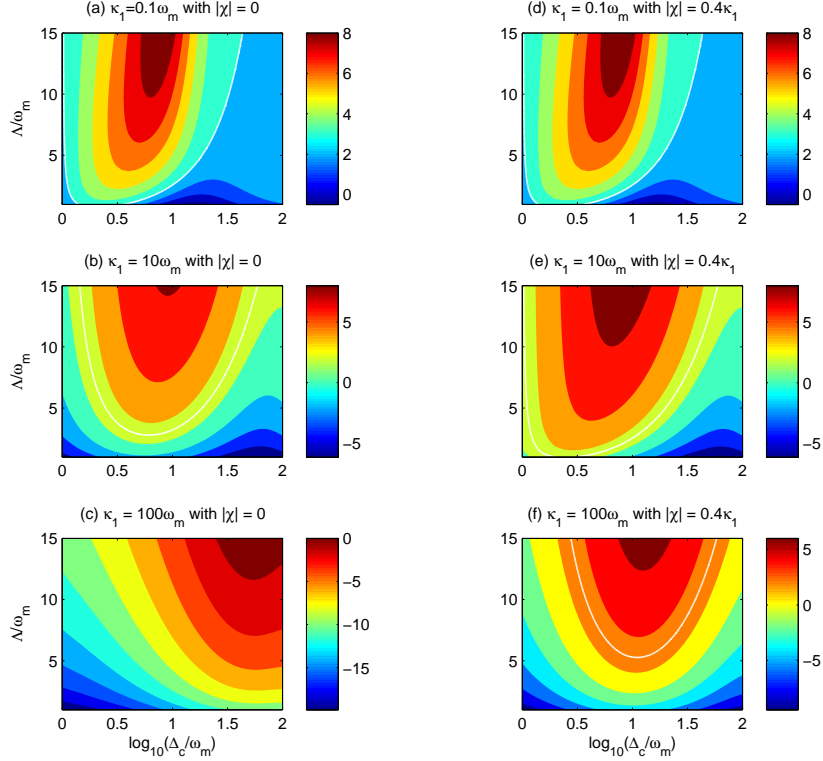}}}
\caption{ Optimal mechanical squeezing $S^{\theta = 0}_{dB}$ of the mechanical resonator (in units of dB) versus $\Lambda$ and $\Delta_c$ for different values of
$\kappa_1$ and $|\chi|$. The white solid lines correspond to mechanical squeezing at 3dB.
Other parameter values are $\gamma_m = 10^{-6}\omega_m$, $n_{th} = 0$, $G_1 = 0.1\omega_m$, $\phi = 0.5\pi$, and $\varphi_b = n\pi$.
} \label{fig5}
\end{figure}

We now turn to investigate the mechanical squeezing which is defined as (in units of dB) \cite{ZhangRong2019}
\begin{eqnarray}
S^{\theta}_{dB} &\equiv& -10\log_{10}(\langle \Delta Z_{\theta, b}^2 \rangle/ \langle \Delta X^2 \rangle_{ZPF}) \nonumber\\
 &=& -10\log_{10}(2 \langle \Delta Z_{\theta, b}^2 \rangle),
\end{eqnarray}
with $\langle \Delta X^2 \rangle_{ZPF} = 0.5$ being the zero-point fluctuations and $Z_{\theta, b} = X_b \cos{\theta} + Y_b \sin{\theta}$. Here, $S^{\theta}_{dB} = 0$ when the fluctuations of the mechanical resonator $\langle \Delta Z_{\theta, b}^2 \rangle$ is equal to the zero-point fluctuations $\langle \Delta X^2 \rangle_{ZPF}$. Note that $S^{\theta}_{dB}$ is the squeezing of quadrature $X_b$ ($Y_b$) when $\theta = 0$ ($\theta = \pm \pi/2$).
The mechanical squeezing can be calculated by using Eq.(\ref{dV}) of the previous section and the above definition.

In Fig.4, we plot the mechanical squeezing as functions of $\theta$ for different values of $|\chi|$ and $\kappa_1$. In the absence of the nonlinear interaction between two cavity modes ($|\chi| = 0$), the maximal values of mechanical squeezing $S^{\theta}_{dB}$ are larger than 3dB if $\kappa_1 = 0.1\omega_m$ while the maximal values of $S^{\theta}_{dB}$ are negative if $\kappa_1 = 100\omega_m$. 
In the presence of the nonlinear interaction between two cavity modes, 
the maximal values of $S^{\theta}_{dB}$ can be larger than 3dB both for 
$\kappa_1 = 0.1\omega_m$ and $\kappa_1 = 100\omega_m$. Particularly, $S^{\theta}_{dB}$ is maximal when $\theta = 0, \pm \pi$ for all cases. In the following, we will plot the maximal values of the mechanical squeezing with $\theta = 0$.

In the present work, the parameter $\Lambda$ is defined as $\Lambda = 3\eta(4|\beta|^2\cos^2{\varphi_b} + 1)$.
A nonlinear amplitude of $\eta = 10^{-4}\omega_m$ can be achieved.
Given $\eta$ and $|\beta|$, the maximal value of the parameter $\Lambda$ is $\Lambda_{max} = 3\eta(4|\beta|^2 + 1)$ when $\varphi_b = n\pi$ with
$n = 0,1,2,...$. In Fig.\ref{fig5}, we plot the optimal mechanical squeezing $S^{\theta = 0}_{dB}$ of the mechanical resonator (in units of dB) versus $\Lambda$ and $\Delta_c$ for different values of
$\kappa_1$ and $|\chi|$ with $\varphi_b = n\pi$. From Figs.\ref{fig5}(d)-\ref{fig5}(f), we find the optimal mechanical squeezing $S^{\theta = 0}_{dB}$ can be larger than 3dB when $\varphi_b = n\pi$. In practice, the Duffing nonlinearity, mechanical frequency, and optomechanical coupling strength should be tuned carefully in order to make sure $\varphi_b = n\pi$. Fortunately, the optimal mechanical squeezing can still be larger than 3dB even when $\varphi_b \neq n\pi$.
For example, in the case of $\eta = 10^{-4}\omega_m$, $|\beta| = 111$, and $\varphi_b = \pi/6$, the parameter $\Lambda$ is about $11\omega_m$ and the 3dB limit can be overcome when 
$\varphi_b = \pi/6$ as one can see from Figs.\ref{fig5}(d)-\ref{fig5}(f) clearly.
In fact, the 3dB limit can be surpassed for a wide range of $\varphi_b$ since $|\beta|$ can be larger than 500 as one can see from Fig.2(c) and $\Lambda$ could be several times larger than the mechanical frequency $\omega_m$.

In Figs.\ref{fig5}(a)-\ref{fig5}(c), the second-order nonlinear medium is not put into the cavity. From Fig.\ref{fig5}(a), one can see the optimal mechanical squeezing can be larger than 3dB if the Duffing nonlinearity is strong enough in the resolved sideband regime in the absence of second-order nonlinear medium $\chi^{(2)}$.
This is consistent with the results of \cite{Lv2015}.
The optimal mechanical squeezing decreases with the increase of the decay rate of cavity $\kappa_1$.
For instance, if the decay rate of the cavity is much larger than the frequency of the mechanical resonator ($\kappa_1 = 100\omega_m$), then
the optimal mechanical squeezing of the mechanical resonator could not be larger than 3dB (see Fig.\ref{fig5}(c)).
If the second-order nonlinear medium is put into the cavity, then the optimal mechanical squeezing overcomes the 3dB limit
even in the HURSB regime as one can see
from Fig.\ref{fig5}(f). This could be explained as follows.
The elements $A_{11} = |\chi|\cos{(2\phi)} - \kappa_1/2$
and $A_{22} = -|\chi|\cos{(2\phi)} - \kappa_1/2$ of the drift matrix $A$ in Eq.(\ref{A}) play import roles in the dynamics of $X_{a_1}$ and $Y_{a_1}$ as we can clearly see from Eq.(\ref{dfdt}). Therefor, the nonlinear interaction between two cavity modes can adjust the dynamics of $X_{a_1}$ and $Y_{a_1}$. The decay rate of the cavity mode $a_1$ could be decreased significantly and sideband resolution is improved. 
As a result, the optimal mechanical squeezing can overcome the 3dB limit in the HURSB regime when there is nonlinear interaction between the cavity mode $a_1$ and the fast-decay mode $a_2$.

In the absence of the second-order nonlinear medium with $|\chi| = 0$, the mechanical
squeezing depends heavily on the decay rate $\kappa_1$, i.e., it decreases with the increase of the decay
rate $\kappa_1$ significantly (Figs.\ref{fig5}(b)-\ref{fig5}(c)).
However, if we put the second-order nonlinear medium into the cavity, the optimal mechanical squeezing is insensitive to the
decay rate $\kappa_1$ as one can find in Figs.\ref{fig5}(e)-\ref{fig5}(f).

Note that we have assumed $\phi = 0.5 \pi$ in Figs.5(a)-5(f). In fact, the parameter $\phi$ plays a crucial role in the generation of mechanical squeezing.
We have also
calculated the optimal mechanical squeezing $S^{\theta = 0}_{dB}$ carefully for different values of $\phi$ (not shown here). It is found that the second-order nonlinear medium decreases the optimal mechanical squeezing $S^{\theta = 0}_{dB}$ both in the resolved sideband regime and HURSB regime
when $ 0 \lesssim \phi \lesssim 0.3 \pi$. Thus, the parameter $\phi$ must be larger than $0.3\pi$ in order to generate optimal mechanical squeezing larger than 3dB in the HURSB regime. This is a restriction on the choice of the realistic parameters.

\subsection{Robustness against thermal fluctuations of mechanical mode}

\begin{figure}[tbp]
\centering {\scalebox{0.6}[0.6]{\includegraphics{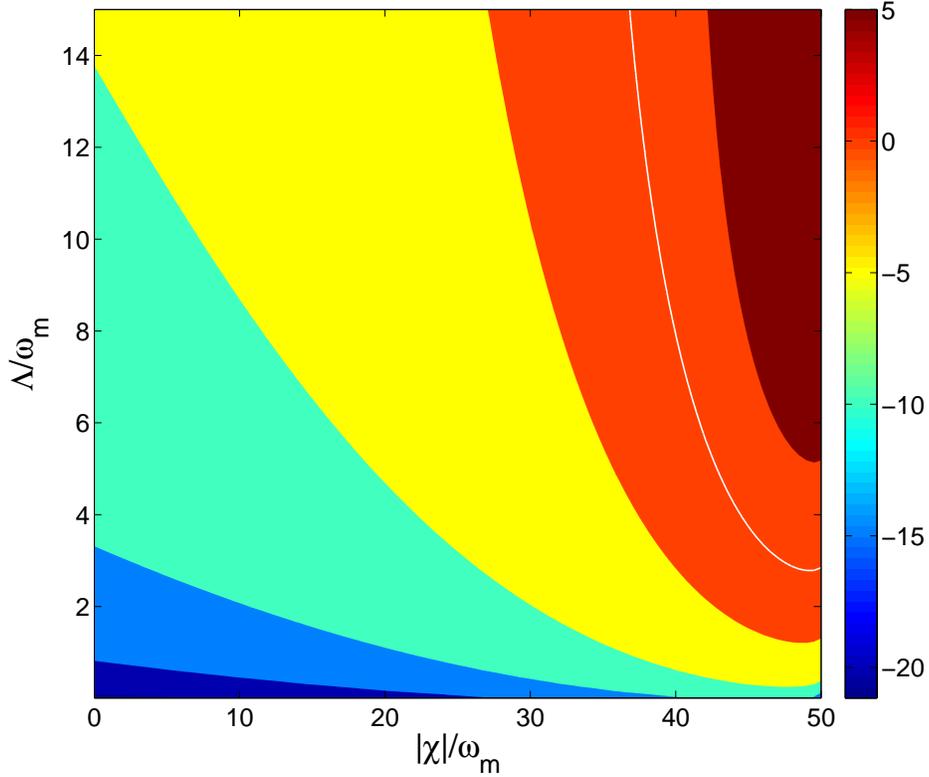}}}
\caption{Optimal mechanical squeezing $S^{\theta = 0}_{dB}$ of the mechanical resonator (in units of dB) versus $\Lambda$ and $|\chi|$.
The white solid line corresponds to mechanical squeezing at 3dB.
Other parameter values are $\gamma_m = 10^{-6}\omega_m$, $n_{th} = 1000$, $G_1 = 1.5\omega_m$,
$\Delta_c = 10\omega_m$, $\kappa_1 = 100\omega_m$, $\phi = 0.5\pi$, and $\varphi_b = n\pi$.
} \label{fig6}
\end{figure}

Now, we investigate the influence of the thermal fluctuations on the optimal mechanical squeezing. Here, we assume the frequency of the mechanical resonator is $\omega_m = 2\pi \times 20 MHz$. Therefore, the thermal occupation $n_{th}$ can be calculated from the relation $n_{th} = 1/\{\exp{[(\hbar \omega_m)/(k_b T)]} - 1\}$ where $k_b$ is the Boltzmann constant. The thermal occupations are 500, 1000, and 5000 for 
$T \approx 0.5 K$, $T \approx 1 K$, and $T \approx 5 K$, respectively.

In Fig.\ref{fig6}, we plot the optimal mechanical squeezing of the mechanical resonator versus $\Lambda$ and $|\chi|$
with $\gamma_m/\omega_m = 10^{-6}$, $n_{th} = 1000$, $G_1 = 1.5\omega_m$,
$\Delta_c = 10\omega_m$, $\kappa_1 = 100\omega_m$, $\phi = 0.5\pi$, and $\varphi_b = n\pi$.
If there is no Duffing nonlinearity or second-order nonlinear medium, large mechanical squeezing can not be achieved.
However, if the second-order nonlinear medium and Duffing nonlinearity are
chosen appropriately the optimal mechanical squeezing can overcome the 3dB limit
even in the HURSB regime and in the presence of thermal fluctuation of the mechanical mode with $n_{th} = 1000$.
This shows that the large and robust mechanical squeezing is a combined effect of the Duffing nonlinearity
of the mechanical mode and the second-order nonlinearity medium $\chi^{(2)}$ in the cavity.

\begin{figure}[tbp]
\centering {\scalebox{0.6}[0.6]{\includegraphics{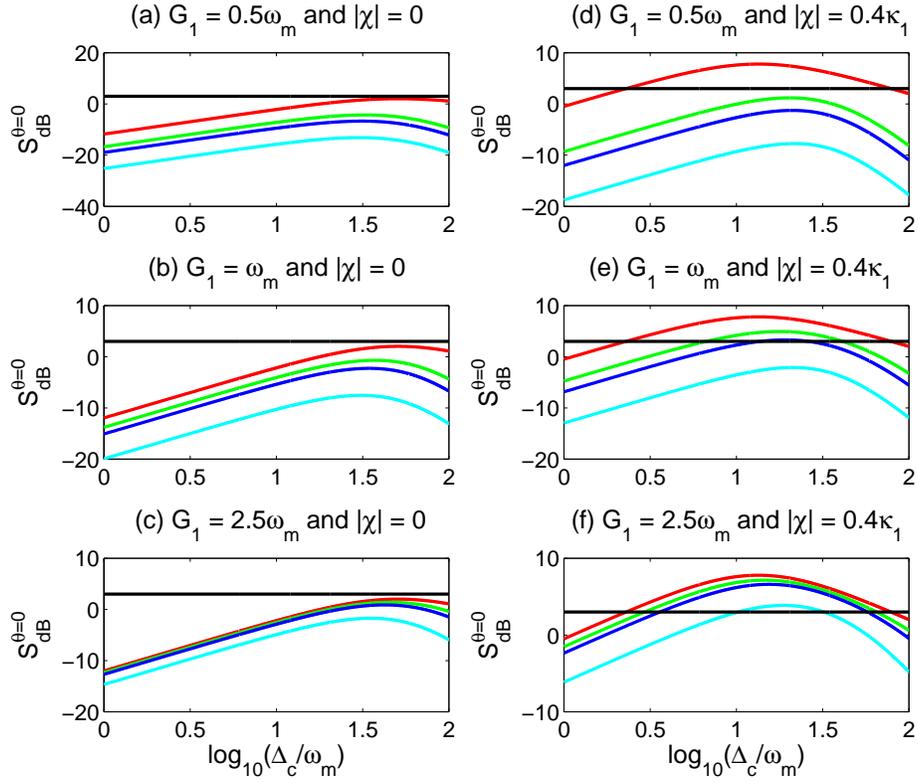}}}
\caption{ Optimal mechanical squeezing $S^{\theta = 0}_{dB}$ of the mechanical resonator (in units of dB) versus $\Delta_c$ for different values of
$G_1$ and $|\chi|$. The black solid lines correspond to mechanical squeezing at 3dB.
The red, green, blue, and cyan lines correspond to $n_{th} = 0$, $n_{th} = 500$, $n_{th} = 1000$, and $n_{th} = 5000$, respectively.
Other parameter values are $\gamma_m = 10^{-6}\omega_m$, $\Lambda = 8\omega_m$, $\phi = 0.5\pi$, and $\varphi_b = n\pi$.
} \label{fig7}
\end{figure}

In order to show the influence of the thermal fluctuations on the mechanical squeezing more clearly, we plot
the optimal mechanical squeezing (in units of dB) as functions of $\Delta_c$ for different values of $G_1$ and $|\chi|$
with $n_{th} = 0$ (red lines), $n_{th} = 500$ (green lines), $n_{th} = 1000$ (blue lines), and $n_{th} = 5000$ (cyan lines)
in Fig.\ref{fig7}.
It is easy to see that in the absence of the second-order nonlinear medium ($|\chi| = 0$),
the 3dB limit can not been overcome even for deep strong coupling regime with $G_1 = 2.5\omega_m$ as
one can find from Figs.\ref{fig7}(a)-\ref{fig7}(c).
However, if the second-order nonlinear medium $\chi^{(2)}$ is considered, the situation is very different.
In the case of $G_1 = 0.5\omega_m$ and $|\chi| >0$ of Fig.\ref{fig7}(d),
the optimal mechanical squeezing can be larger than 3dB for $n_{th} = 0$ and $n_{th} = 500$
while the 3dB limit can not been overcome for $n_{th} = 1000$ and $n_{th} = 5000$.
If we increase the coupling strength between the optical and mechanical modes $G_1$, then the 3dB limit can be surpassed even at the high temperature $10K$ with the corresponding thermal occupation $n_{th} = 5000$ as one can see from the cyan line in Fig.7(f).

\section{Conclusion}
In the present work, we have proposed an efficient scheme to generate large and robust mechanical squeezing
beyond the 3dB limit in the HURSB regime for realistic parameters.
The system was formed by a standard optomechanical system with a second-order nonlinear medium $\chi^{(2)}$ in a cavity and
the Duffing nonlinearity of the mechanical mode. In fact, a strong Duffing nonlinearity could be achieved by coupling the
mechanical mode to an auxiliary system as point out in \cite{Lv2015}. There are two modes in the cavity.
One is the fundamental mode. The other is the second-order mode.
We assumed the decay rate of the second-order mode is very large.
In the adiabatic approximation, we derived an effective quantum Langevin equations of the model.
The influence of the second-order nonlinear medium $\chi^{(2)}$ and Duffing nonlinearity
on the mechanical squeezing was discussed carefully.

In the absence of the second-order nonlinear medium $\chi^{(2)}$, the mechanical squeezing $S_{dB}$ decreases with the
increase of the decay rate of the cavity significantly and it could be negative for HURSB regime.
However, if we put the second-order nonlinear medium into the cavity, the mechanical squeezing is insensitive with the
decay rate of the cavity and $S_{dB}$ can be larger than 3dB even when the decay rate of the cavity is much larger than the
frequency of the mechanical resonator.

Then, we discussed the influence of the thermal fluctuations of the mechanical mode on the mechanical squeezing in the HURSB regime.
On the one hand, the mechanical squeezing can not be larger than 3dB without the second-order nonlinear medium if the thermal fluctuations
of the mechanical mode is considered. On the other hand, the mechanical squeezing could be larger than 3dB even at high temperature when
the second-order nonlinear medium is put into the cavity. Thus, we have shown that large and robust mechanical squeezing
beyond the 3dB limit can be generated in the HURSB regime.

\section*{Funding}
This research was supported by Zhejiang Provincial Natural Science Foundation of China under Grant No. LZ20A040002; National Natural Science Foundation of China (11047115, 11775190, 11365009); 
Scientific Research Foundation of Jiangxi (20122BAB212008 and 20151BAB202020.)

\section*{Acknowledgments}
We thank two anonymous referees for their useful comments and suggestions. 
J. S. Z thanks Wen-Jie Nie for valuable discussions. 

\section*{Disclosures}
The authors declare no conflicts of interest.

\bibliography{ref}

\end{document}